\documentclass[11pt,reqno]{article}
\usepackage{amsmath}
\usepackage{amsfonts}
\usepackage{amssymb}
\usepackage{latexsym}
\usepackage[dvips]{graphicx}
\usepackage{epsf}
\usepackage{bbm}

\allowdisplaybreaks \numberwithin{equation}{section}

\newcommand{\be}{\begin{equation}}
\newcommand{\ee}{\end{equation}}
\newcommand{\bea}{\begin{eqnarray}}
\newcommand{\eea}{\end{eqnarray}}
\newcommand{\nn}{\nonumber}

\let\a=\alpha \let\b=\beta  \let\g=\gamma  \let\d=\delta
     \let\th=\theta  \let\k=\kappa \let\l=\lambda
\let\m=\mu    \let\n=\nu          \let\r=\rho \let\om=\omega
\let\s=\sigma \let\t=\tau     
    
\let\G=\Gamma    \let\L=\Lambda 
     
  \let\eps=\epsilon

\let\==\equiv

\def\FF{{\cal F}} 
 
\def\LL{{\cal L}}

\newcommand{\f}{\frac}

\newcommand{\Id}{{\mathbbm 1}}
\newcommand{\p}{\partial}
\newcommand{\na}{\nabla}
\newcommand{\Tr}{{\rm Tr}}
\newcommand{\tr}{{\rm tr}}
\newcommand{\w}{\wedge}

\newcommand{\omb}{\bar{\om}}

\newcommand{\cb}{\bar{c}}

\newcommand{\psib}{\overline{\psi}}

\newcommand{\wt}{\tilde{w}}

\newcommand{\Qh}{\hat{Q}} 
\newcommand{\Fh}{\hat{F}}
\newcommand{\Sh}{\hat{S}}

\newcommand{\wh}{\hat{w}}
\newcommand{\ws}{\check{w}}
\newcommand{\wb}{\bar{w}}
\newcommand{\Ph}{\hat{P}}
\newcommand{\Ps}{\check{P}}
\newcommand{\Pb}{\bar{P}}

\newcommand{\la}{\langle}
\newcommand{\ra}{\rangle}

\newcommand{\Ref}[1]{(\ref{#1})}

\begin{document}

\thispagestyle{empty}

\begin{center}
 {\LARGE\bfseries Perturbative quantum gravity \\[3mm] with the Immirzi parameter}
\\[10mm]
{\large Dario Benedetti${}^a$ and Simone Speziale${}^b$} \\[3mm]
{\small\slshape
${}^a$Max Planck Institute for Gravitational Physics (Albert Einstein Institute), \\
Am M\"{u}hlenberg 1, D-14476 Potsdam, Germany\\
{\small ${}^b$Centre de Physique Th\'eorique,\footnote{Unit\'e Mixte de Recherche (UMR 6207) du CNRS et des Universites Aix-Marseille I, Aix-Marseille II et du Sud Toulon-Var. Laboratoire affili\'e \`a la FRUMAM (FR 2291).} CNRS-Luminy Case 907, 13288 Marseille Cedex 09, France}
} 

\end{center}
\vspace{5mm}

\hrule\bigskip

\centerline{\bfseries Abstract} \medskip
\noindent
We study perturbative quantum gravity in the first-order tetrad formalism.
The lowest order action corresponds to Einstein-Cartan plus a parity-odd term, and is known in the literature as the Holst action. The coupling constant of the parity-odd term can be identified with the Immirzi parameter $\g$ of loop quantum gravity. We compute the quantum effective action in the one-loop expansion. As in the metric second-order formulation, we find that in the case of pure gravity the theory is on-shell finite, and the running of Newton's constant and the Immirzi parameter is inessential.
In the presence of fermions, the situation changes in two fundamental aspects. First, non-renormalizable logarithmic divergences appear, as usual. 
Second, the Immirzi parameter becomes a priori observable, and we find that it is renormalized by a four-fermion interaction generated by radiative corrections. We compute its beta function and discuss possible implications. The sign of the beta function depends on whether the Immirzi parameter is larger or smaller than one in absolute value, and $\g^2=1$ is a UV fixed-point (we work in Euclidean signature). Finally, we find that the Holst action is stable with respect to radiative corrections in the case of minimal coupling, up to higher order non-renormalizable interactions.
\bigskip
\hrule\bigskip

\section{Introduction}

The coupling constants of general relativity are the cosmological constant, $\L$, and Newton's constant, $G$. They measure respectively the resistence of spacetime to expansion and bending. A simple extension of the theory is to take the metric and the connection as independent fields, which allows the spacetime manifold to have non-trivial torsion. In this framework, largely unconstrained by current observations \cite{Hehl,Shapiro,Trautman}, there is a new fundamental coupling constant. It has the same dimensions of Newton's constant, and it can be conveniently parametrized as $G\g$, with $\g$ a real dimensionless coupling. This additional constant enters the coupling of the gravitational field to the elementary spin of particles sourcing the torsion.

From a lagrangian perspective, this coupling constant multiplies the term 
$\eps^{\mu\nu\rho\s} F_{\mu\nu\rho\s}$, where $\eps^{\mu\nu\rho\s}$ is the Levi-Civita pseudotensor and $F$ is the curvature of the independent connection.
When torsion is zero, i.e. when the connection is fixed to be the metric-dependent Levi-Civita connection, its curvature coincides with the Riemann tensor, and the above term vanishes identically due to the first Bianchi identities.
For this reason, $\eps^{\mu\nu\rho\s} F_{\mu\nu\rho\s}$ is usually discarded in both the classical and the quantum theories.
This term has been known for a long time \cite{Weinberg,Hojman}, and it acquired much attention in the last twenty years with the development of loop quantum gravity (LQG) \cite{LQG}. As shown by Holst \cite{Holst}, the term is necessary to perform the canonical transformation from the traditional Arnowitt-Deser-Misner variables for general relativity in the Hamiltonian formalism (spatial metric and extrinsic curvature), to the gauge-theory-like Ashtekar-Barbero variables \cite{Ashtekar,Barbero} used in LQG. Furthermore, the coupling constant $\gamma$ coincides with the Immirzi parameter, that is the real number parametrizing the canonical transformation \cite{Immirzi}.
Conforming with this literature, we will refer to the term $\eps^{\mu\nu\rho\s} F_{\mu\nu\rho\s}$ as the Holst term, and to its coupling constant $\g$ as the Immirzi parameter.

The classical irrelevance of the Holst term can be formally extended at the quantum level: if one restricts the integration to invertible metrics, then the integral over the connection can be performed exactly, and the usual second order formalism is recovered, with no left-over dependence on $\gamma$. 
On the other hand, if one allows degenerate metrics in the path integral, then configurations with non-zero torsion will contribute. The idea that a phase with $\la g_{\m\n} \ra =0$ plays a role in non-perturbative quantum gravity has often appeared in the literature (e.g. \cite{Tseytlin,Smolin:1979uz,Witten,Giddings}).
One might wonder whether such contributions trigger a non-perturbative quantum relevance of $\g$. 
Loosely speaking, one is considering the possibility that the gravitational degrees of freedom are better described at high energy by the connection, rather than the metric.
This is indeed the set-up of LQG. This approach suggests that $\g$ plays a major role in non-perturbative quantum gravity. In particular, the famous kinematical \emph{area gap} of the theory, $A = 4\sqrt{3}\pi \hbar G \g$,  is proportional to it \cite{RovelliThiemann}, and recent definitions of the non-perturbative quantum dynamics depend explicitly on it \cite{EPRL}. A possible running of $\g$ within this framework is discussed for example in \cite{Rivasseau}.
Within a quantum field theory approach, a natural setting in which to look for such a non-perturbative role of $\g$ could perhaps be that of the asymptotic safety scenario \cite{Weinberg79,NiedermaierReuter,Codello:2008vh,Litim,BMS}, and an attempt of contact in this direction has been recently considered in \cite{DaumReuter}, using a specific truncation of the functional renormalization group equations.

Motivated by these ideas, one wonders whether the relevance of this coupling constant at the quantum level 
can be understood using more conventional methods.
In this paper we consider perturbative quantum gravity, and compute the 1-loop effective action of gravity with the Immirzi parameter. Although non-renormalizable, the theory makes sense as an effective field theory \cite{Donoghue}. 
We use standard methods, such as the background field formalism and the heat kernel expansion, which have been extensively applied to quantum gravity \cite{BuchbinderOdintsovShapiro}.
The novelties of our work are the use of the tetrad and connection as independent variables, and the inclusion of parity-odd terms.

Like in the second-order formulation \cite{'tHooft,ChristensenDuff}, we find that the pure gravity theory is 1-loop renormalizable, and a running of the coupling constant $G$ is inessential. In addition, we find that also the running of $\g$ is inessential. 
Motivated by making contact with non perturbative results, we also consider an off-shell renormalization condition for $\g$. In this scheme, both $G$ and $\g$ run, and we find two fixed points, at $\g=0$ and $\g=\infty$, respectively IR and UV attractive. 

The situation changes radically if sources of torsion are present, notably in the case of fermion coupling to gravity. 
Now the first and second order formulations differ by the presence, in the first case, of an extra four-fermion contact interaction, and can be thus distinguished as physical theories. The contact interaction is induced by the coupling between the connection and the fermionic currents, and depends explicitly on the Immirzi parameter, which then becomes a quantity a priori classically measurable. 

At the quantum level, we find that $\g$ becomes an essential parameter, and furthermore that it is naturally renormalized, since four-fermion interactions are generated by radiative corrections. 
These radiative corrections are found also in the second-order formulation of the theory \cite{DeserNieuwenhuizen,Barvinsky:1981rw}, where they belong to the non-renormalizable type. 
On the contrary, our results show that working in the first-order formalism, the Holst term provides a natural counter-term to them. 
We focus mainly on the simplest gravity-fermion system, that is minimally-coupled Majorana spinors. In this case, we can use previous results \cite{Barvinsky:1981rw} to derive the 1-loop effective action for the coupled system.
We compute the beta function of the Immirzi parameter, and show that in the presence of fermions there are no fixed points, apart from the special values $\g^2=1$ which correspond to general relativity in self-dual variables \cite{Ashtekar,Plebanski}.
Non-renormalizable logarithmic divergences also appear, as expected from the second order formalism \cite{DeserNieuwenhuizen,Barvinsky:1981rw}. Finally, we comment on implications and extensions to non-minimal couplings, which are likely to require a more general bare gravitational action.

In all considered cases, in the absence of a cosmological constant the running of the Immirzi parameter is driven by quadratic divergences. These are usually discarded in the framework of dimensional regularization, but if the cut-off provides an actual physical scale in the effective field theory, then it would be uncautious to neglect them. Quadratic divergences might play important roles in systems coupled to gravity, as emphasized in \cite{Robinson,Toms} (see however \cite{Anber}).

The paper is organized as follows. In Sec.~\ref{Sec:Action} we introduce the general first-order formalism. We review the Holst action and its relation to general relativity and theories with torsion. In Sec.~\ref{Sec:1loop} we present the general 1-loop algorithm, and then the details of the pure gravity calculation in Sec.~ \ref{Sec:Perturb}, Sec.~\ref{Sec:HKexp} and Sec.~\ref{Sec:Renorm}. In Sec.~\ref{Sec:Fermi} we study the fermion-coupled system.
Conclusions and perspectives are collected in Sec.~\ref{Sec:Conclusions}. 
We work in Euclidean signature, as customary in the perturbative quantum gravity literature. The appendix contains a list of conventions and useful formulas.

\section{First-order action for gravity}
\label{Sec:Action}

As we will later couple gravity to fermions, we work from the beginning with tetrads instead of the metric.
In the first order formalism we take as independent variables the tetrad $e_\mu^I$, and a connection $\om_\mu^{IJ}$ in the local gauge group SO(4). 
If we do not require a priori the invertibility of the tetrads, there are only six possible invariants under diffeomorphisms and local gauge transformations. Accordingly, the most general action reads
\bea \label{S1}
&& S[e,\om] = \f{1}{\k^2} \left\{ 2\L \int e  - \int \f12 \eps_{IJKL} e^I\w e^J\w F^{KL} 
- \f1\g \int e_I\w e_J\w F^{IJ}\right\} \\ \nn
&& \qquad + \a_1 \int \eps_{IJKL} F^{IJ} \w F^{KL}
+ \a_2 \int F^{IJ} \w F_{IJ} + 
\a_3 \int \Big(T^I\w T_I-e_I\w e_J\w F^{IJ}\Big),
\eea
where $F^{IJ}(\om)=d\om^{IJ} +\om^I{}_K\w \om^{KJ}$ is the curvature, and $T^I=d_\om e^I$ the torsion.
The first two terms give the Einstein-Cartan action \cite{Hehl,Utiyama,Kibble,Sciama}, with $\Lambda$ the cosmological constant and $\k^2=16\pi G$, and the third is the Holst term $\eps^{\mu\nu\rho\s} F_{\mu\nu\rho\s}$. 
The remaining three terms, in the second line of \eqref{S1}, are the topological Euler, Pontrjagin and Nieh-Yan classes.
$T^I \w T_I$ is the only torsion-squared term that can be written without the inverse metric, and it coincides with the Holst term up to a boundary contribution (the Nieh-Yan invariant).
The Holst, Pontrjagin and Nieh-Yan terms are parity-odd in vacuum.

When the Nieh-Yan invariant vanishes, the coupling constant $\g$ coincides with the Immirzi parameter which permits to write the canonical theory as an SU(2) gauge theory \cite{Holst}. When the Nieh-Yan invariant is non-zero, also $\alpha_3$ contributes to this canonical transformation \cite{Date}. In the following, we will consider only spacetimes with trivial topologies, thus we can identify the sole coupling $\g$ with the Immirzi parameter.\footnote{We would like to stress that from this perspective, the Immirzi parameter is simply a coupling constant which is necessarily present in the action if one chooses to work with independent tetrad and connection. For instance, it appears also in the same first-order formulation applied to supergravity \cite{Kaul}.}

Equivalence of \Ref{S1} with general relativity is easily established in the sector of invertible tetrads. 
The field equations are
\bea\label{eq1}
&& \left(\Id/\g + \star\right)^{IJ}_{KL} e^K \wedge d_\omega e^L = 0,
\\ \label{eq2}
&& \left(\Id/\g + \star\right)^{IJ}_{KL} e_J \wedge F^{KL}(\om) - \f\L6 \eps^{I}{}_{JKL} e^J\w e^K\w e^L = 0.
\eea
Here $\star=(1/2)\eps^{IJ}_{KL}$ and $\Id=\d^{IJ}_{KL}=\d^I_{[K} \d^J_{L]}$, and the square brackets mean weighted antisymmetrization. Notice that $\Id/\g + \star$ is not invertible for $\g^2=1$, where it becomes a projector on the self/antiself-dual parts of the Lorentz group. Assuming $\g^2\neq 1$, and an invertible tetrad, \Ref{eq1} implies the vanishing of the torsion, and it has the unique solution
\be\label{oe}
\om_\m^{IJ}(e) = e^I_\n \nabla_\mu e^{\nu J},
\ee
where $\na_\m$ denotes the covariant derivative with Levi-Civita connection. Then its curvature gives the Riemann tensor,
$F^{IJ}_{\m\n}(\om(e)) \equiv e^{I\r} e^{J\s} R_{\r\s\m\n}(e)$, and \Ref{eq2} reduces to 
the Einstein's equations in the tetrad form $R^\m_I -\f12 R e^\m_I + \L e^\m_I = 0$. 
Notice that the term proportional to the Immirzi parameter vanishes due to the Bianchi identity $\eps^{\mu\nu\rho\s} R_{\mu\nu\rho\l}(e)\equiv 0$. 
Hence we recover the standard metric formulation, and the Immirzi parameter completely drops out of the theory. 

For the special value $\g=1$ (and similarly for $\g=-1$), the antiself-dual components of the fields completely drop out of the formalism, and one is dealing with a formulation of gravity in terms of self-dual variables only \cite{Ashtekar,Plebanski}.

At a formal level, the Immirzi parameter allows us to ``interpolate'' between different formulations of the theory: for $\g\to\infty$ we obtain the simple Einstein-Cartan version of first-order gravity, whereas for $\g\to 0$ we recover the second-order formalism with no torsion. This can be seen when the Nieh-Yan invariant vanishes. Then the Holst term equals the torsion-squared term, and by introducing an auxiliary 2-form field $B$ we can rewrite it in the path integral as $\f{1}{\g} T\w T \to 2B\w T + \g B\w B$: the limit $\g=0$ now yields a Lagrange multiplier enforcing $T=0$, i.e. the second-order theory.
These changes of formulations have clearly no consequences at the classical level, but might lead to different quantum theories. In this respect, the above formal manipulation has been used in \cite{FS} to argue that $\g$ controls the quantum fluctuations of the vanishing torsion condition.

If we do not require the invertibility of the tetrad, \Ref{S1} is the most general action invariant under diffeomorphisms and local Lorentz transformations: All other invariants require either the invertibility of the tetrad or the use of auxiliary fields.
However, we are not aware of any symmetry or other mechanism to protect this remarkably simple structure of the action.
If one allows the use of inverse tetrads, there is again an infinite number of terms that can appear in the action.
To study the invariants in the sector of invertible tetrads, it is convenient to parametrize the connection as
\be\label{omdec}
\om = \om(e) + K,
\ee
in terms of the spin connection \Ref{oe} and the contorsion tensor $K$, which satisfies $K^I{}_J \w e^J = T^I$. 
The contorsion is antisymmetric on the last two indices, and we can use the tetrad to project it to a spacetime tensor $K_{\mu\n\r}=-K_{\mu\r\n}$.

Using \Ref{omdec}, we can package the invariants in terms of the more familiar
contractions of the Riemann tensor $R_{\m\n\r\s}(e)$, and of the contorsion $K_{\mu\n\r}$. 
The unique invariant of zero dimensions is the usual volume term det $e_\m^I$, already included in \Ref{S1}. 
The dimension-two invariants of \Ref{S1} decompose as follows,
\bea \label{epseeF}
&& \f12 \eps_{IJKL} e^I\w e^J\w F^{KL} = 
(d^4x) e [R(e) + K_{\m\n\r} K^{\n\m\r} - K^\m{}_{\m\r} K_\n{}^{\n\r}] + {\rm boundary \ term}, \\ \label{eeF}
&& e_I\w e_J\w F^{IJ} = (d^4x) K_{\m\n}{}^{\l} K_{\r\s\l} \eps^{\m\n\r\s} + {\rm boundary \ term}.
\eea
Hence, we can view the theory in \eqref{S1} as a theory of gravity plus (non-dynamical) torsion \cite{Shapiro}.
In particular, we see from \Ref{eeF} that the Holst term is non-trivial on-shell only in the presence of torsion, as anticipated above.
The four terms appearing here do not exhaust the set of dimension-two invariants of $R_{\m\n\r\s}(e)$
and $K_{\mu\n\r}$.
The complete list is
\begin{subequations}\label{dim2}\begin{align}
\label{dim2R} & eR(e), \\
\label{dim2Kpari} & eK^{\m\n\r} K_{\m\n\r}, \qquad eK^{\m\n\r} K_{\n\m\r}, \qquad  eK^{\m}{}_{\m\r} K^\n{}_{\n\r}, \\
\label{dim2Kdis} & K_{\m\n}{}^{\l} K_{\r\s\l} \eps^{\m\n\r\s}, 
\qquad K_{\m\n\r} K^\l{}_{\l\s} \eps^{\m\n\r\s}.
\end{align}\end{subequations}
Two new terms are possible, with respect to those present at this order in \Ref{S1}, thanks to the invertibility of the tetrad.\footnote{In the literature, e.g. \cite{Hojman}, two additional parity-odd terms are reported,
$K_{\m\n}{}^{\l} K_{\l\r\s} \eps^{\m\n\r\s}$ and  $K^{\l}{}_{\m\n} K_{\l\r\s} \eps^{\m\n\r\s}$. 
However, these are not independent from the two already in \Ref{dim2Kdis}, due to algebraic identities such as 
$\eps_{AB[C}{}^{[E}\d_{ D]}{}^{F]} = - \eps_{CD[A}{}^{[E}\d_{B]}{}^{F]} $.
}

Among the invariants of dimension four, we find the terms $R_{\m\n\r\s}^2$,
$R_{\m\n}^2$, and $R^2$ of the metric formalism, 
as well as many new ones. 
For example, we have a new parity-odd term, the Hirzebruch signature invariant,
$R_{\m\n\r\s} R_{\a\b\g\d} g^{\m[\a} g^{\n\b]} \eps^{\r\s\g\d}$,
related to the Pontryagin class in \Ref{S1}.
We then have terms of type $K^4$, terms of dynamical torsion\footnote{Note that, in presence of $\nabla K \nabla K$ terms, the torsion becomes dynamical and the equivalence between the first- and second-order formulations of gravity gets broken. However, torsion would have a mass of the order of the Planck mass, thus it would still be non-propagating at low energies (see also \cite{Percacci:2009ij} for a nice analogy to the Higgs phenomenon). } $\nabla K \nabla K$, as well as mixed terms, 
such as $K^2 R$, etc. A moment of thought shows that there is a plethora of such terms. For instance, there are 15 independent parity-even $\nabla K \nabla K$ terms \cite{Neville}.

These circumstances make an effective field theory approach appear daunting.
In particular, the question of identifying enough physical observables to distinguish all the coupling constants seems to strongly limit the predictive power of an effective theory based on this formulation. Although this is certainly an interesting set-up to explore, in the following we will mainly concentrate on a bare action of the type \eqref{S1}. In view of the list of invariants in \eqref{dim2}, one might think that, as we are not including the full list of dimension-two terms in our action, radiative corrections will produce non-renormalizable divergences proportional to the missing terms. However, we will see that both for pure gravity, and for gravity minimally coupled to fermions, such divergences are harmless.

\section{1-loop effective action: the algorithm}
\label{Sec:1loop}

The 1-loop effective action for general relativity has been extensively studied in the literature,
however, to the best of our knowledge, never in the set-up here considered, that is the tetrad and connection taken as independent fields, and the inclusion of parity-odd terms.
The closest to it is probably the first-order case of \cite{BuchbinderShapiro}, but it differs for the use of metric variables and, more importantly, for the absence of the Holst term.

To quantize the theory we will use the background-field method (e.g. \cite{BuchbinderOdintsovShapiro,WeinbergV2}) and a 1-loop perturbative expansion. 
We introduce the change of variables
\be\label{pertdef}
e^I_\m \to e^I_\m +\kappa f^I_\m, \qquad
\om^{IJ}_\m \to \om^{IJ}_\m + \kappa w^{IJ}_\m = \omb^{IJ}_\m(e) + K^{IJ}_\m + \kappa w^{IJ}_\m
\ee
with $f$ and $w$ the quantum fields, $e$ and $\om$ the background fields, which we take generic (in particular off-shell) for the time being.
The only restriction we impose on the background tetrad is to be invertible, which allows us to decompose $\om$ as in \Ref{omdec}.
This has important consequences, for as we discussed above, it means that there is an infinite number of invariants that exist. The quantization will then generate an infinite number of terms, and we are led to a situation familiar from the metric formalism.

It would certainly be interesting to consider more general backgrounds. Classical solutions with non-invertible tetrads are known (e.g. \cite{Tseytlin,D'Auria}), and these could introduce interesting physical effects through torsion and parity breaking. However, apart from physical motivations one might wish to discuss, we stress here that there are \emph{technical} necessities to take the background invertible. First of all, the action starts with a cubic term (of the type $ee\om$), hence the Hessian around a non-invertible tetrad would be badly degenerate,\footnote{Even identically zero, if we wanted to do perturbation theory around the vanishing solution $e_\m^I=0$, which proved so succesfull in 2+1 gravity \cite{Witten}.} and we would have troubles in using perturbation theory. A related issue comes from the need to gauge-fix the invariance under diffeomorphism. The usual covariant gauge-fixings require the inverse background metric.\footnote{One could consider alternative gauge-fixing procedures, but this does not seem to improve the situation. One could consider a generic background, and introduce an auxiliary invertible metric only through the gauge-fixing term, as done in topological field theories. The physical results should not depend on the auxiliary metric, but the intermediate calculations will again introduce all possible invariants in the auxiliary, invertible metric. 
Or, one could take a non-covariant gauge-fixing such as $h_{\m0}=0$ (e.g. \cite{Mattei}), but then also non-invariant expressions should be produced. }

Before entering the details of the perturbative expansion, let us give a brief overview of the algorithm.
The 1-loop effective action is determined by the Hessian of $S$ evaluated on the background fields.
On-shell, the Hessian has zero-modes because of gauge and diffeomorphism symmetries. Therefore, we need to add a gauge-fixing term, as well as a ghost term to represent the Faddeev-Popov determinant associated to the gauge fixing.
We define the total action
\be\label{Stot}
S_{\rm tot}[e,\om;f,w;{\rm ghosts}] = S[e+\k f,\om+\k w] + S_{gf}[e,\om;f] + S_{gh}[e,\om;{\rm ghosts}] = \Sh + S_{gh},
\ee
from which we wish to compute the 1-loop effective action
\be\label{Gamma}
\G[e,\om] = S[e,\om] + \f{1}{2} \Tr \ln \hat H - \Tr \ln H_{gh},
\ee
where $\hat H$ and $H_{gh}$ are the Hessians of the gauge-fixed action $\Sh$ and the ghost term $S_{gh}$ respectively, both evaluated on the background fields, 
and Tr denotes the trace of an infinite-dimensional operator (that is, it includes the spacetime integral).
Note that following the background field method's protocol, we are identifying in \eqref{Gamma} the background fields with the mean fields, arguments of the effective action.
At the same time the gauge-fixing term is chosen in such a way to preserve the symmetry under simultaneous transformations of the background and fluctuation fields, while breaking of course the genuine gauge transformations of the fluctuation fields. In this way, one obtains a gauge-invariant effective action, i.e. we are guaranteed that only gauge-invariant terms are generated upon quantization.

The operator traces can be evaluated following the standard trick of rewriting
\be
\Tr \ln H = - \int_0^{\infty} \frac{dt}{t} \Tr \left[ e^{-t H} \right],
\ee
and then using the heat-kernel expansion
\be
\Tr \left[ e^{-t H} \right] = \frac{1}{(4\pi t)^2} 
\int d^4x \sqrt{g} \left[ \tr a_0 + t \, \tr a_1 + t^2 \, \tr a_2 + o(t^3) \right].
\ee
Here tr denotes the trace over the spacetime/internal indices.
Formulas for the heat-kernel coefficients $a_i$ for several type of operators can be found in the literature, 
and we will report below those of our interest. 

In the following, we will be interested in the ultraviolet divergences, 
and it is easily seen that these arise from the first three terms of the heat kernel expansion, after integration over $t$. The remaining terms are UV-finite. 
In a standard fashion, we introduce a UV-cutoff $\L_{UV}$ by substituting the lower bound of integration on $t$ with $1/\L_{UV}^2$. This gives the following regularized expression for the divergent 1-loop contributions to \Ref{Gamma},
\be\label{Gammadiv}
\begin{split}
\G^{\rm div}_{\rm 1-loop} = & -\frac{1}{32 \pi^2} \int d^4x \sqrt{g} \left[ \f12 \L_{UV}^4 \tr \hat{a}_0 + \L_{UV}^2 \tr \hat{a}_1 + 2 \ln(\L_{UV}/\m) \, \tr \hat{a}_2 \right]\\
& + \frac{1}{16 \pi^2} \int d^4x \sqrt{g} \left[ \f12 \L_{UV}^4 \tr a^{gh}_0 + \L_{UV}^2 \tr a^{gh}_1 + 2 \ln(\L_{UV}/\m) \, \tr a^{gh}_2  \right].
\end{split}
\ee
In general, the coefficients contain all possible invariants of the tetrad and contorsion, thus we expect quartic divergences proportional to the volume, and quadratic ones proportional to the various terms in \Ref{dim2}.
In particular, the quadratic ones can renormalize the bare couplings of \Ref{S1}.
Before giving the details of this expression and evaluating the coefficients, we need to discuss the gauge-fixing procedure and the expansion of the action.

\section{Perturbative expansion}
\label{Sec:Perturb}
\subsection{Gauge-fixing and ghosts}
In Einstein-Cartan theory, it is sufficient to gauge-fix the tetrad to remove the degeneracy of the action under diffeomorphisms and Lorentz transformations. This is still true in the presence of the Holst term, provided $\g^2\neq1$. For these special values, the action reduces to general relativity in self-dual variables, and as explained earlier only half the components of the connection can be solved for. Consequently, our formulas will present singularities at $\g^2=1$. In the following we will assume $\g^2\neq 1$ and not discuss general relativity in self-dual variables.

The action is invariant under diffeomorphism and internal gauge transformations, respectively (at the order we are interested in)
\begin{align} \label{diffeo}
& \d_\xi f_\m^I = e_\r^I \na_\m \xi^\r - \xi^\r \om_\r^{IJ} e_{\m J},
\\ & \label{lorentz}
\d_\l f_\m^I = \l^I_J e_\m^J,
\end{align}
with the background fields kept fixed. 
To fix the gauge, we partially follow \cite{DeserNieuwenhuizen}. 
We use the projections $f_{\m\n}=f_\m^I e_{\n I}$, and decompose them into the symmetric, 
$s_{\m\n} = f_{\m\n} + f_{\n\m}$ with trace $s$,
and antisymmetric parts, $a_{\m\n} = f_{\m\n} - f_{\n\m}$.
They transform respectively as
\begin{subequations}\label{gauge}\begin{align}
& \d_\xi s_{\m\n} = \na_\m \xi_\n + \na_\n\xi_\m, 
&& \d_\xi a_{\m\n} = \na_\m \xi_\n - \na_\n\xi_\m + \xi^\r \om_\r^{IJ} e_{\n J} e_{\m I} - \xi^\r \om_\r^{IJ} e_{\m J} e_{\n I}, \\
& \d_\l s_{\m\n} = 0, && \d_\l a_{\m\n} = 2 \l_{IJ} e_\m^J e_\n^I.
\end{align}\end{subequations}

This decomposition has the advantage of disentagling the symmetries. Gauge-fixing terms
can be added to the action as one-parameter families,
\be\label{Sgf}
S_{gf} = \f{1}{2\a} \int e\, \FF^\m\FF_\m + \f{1}{2\b} \int e\, a_{\m\n}a^{\m\n} \, ,
\ee
where
\be
\FF_\m = \na^\n s_{\m\n} - \f{1+\r}{4} \na_\m s,
\ee
and $\a$, $\b$ and $\r$ are real parameters characterizing different gauge choices.
The $\FF$ term is the standard family used to break the diffeomorphism symmetry.
The breaking of the Lorentz symmetry \eqref{lorentz} by the $a_{\m\n}$ term can be seen by a counting argument: $so(4)$ has 6 generators, thus fixing this symmetry means to fix 6 components of $f_{\m\n}$, exactly the number of components of its antisymmetric part $a_{\m\n}$.

Let us write the symmetry transformations \Ref{gauge} as
\bea\nn
s_{\m\n} &\to& s_{\m\n} + \Qh^{(s)}{}_{\m\n}^\a \xi_\a \, ,
\\ \nn a_{\m\n} &\to& a_{\m\n} + \Qh^{(a)}{}_{\m\n}^\a \xi_\a + 2 \l_{\m\n} \, ,
\eea
and the gauge fixing function as $\FF_\m = \Fh_\m^{\r\s} s_{\r\s}$. Then the Faddeev-Popov prescription gives for the ghost term
\be \label{Sgh}
\begin{split}
S_{gh} &= \int e \begin{pmatrix} \cb^\m & \bar\chi^{\m\n} \end{pmatrix} \begin{pmatrix} \Fh_\m^{\a\b} & 0 \\ 0 & \d_{\m\n}^{\a\b} \end{pmatrix} \begin{pmatrix} \Qh^{(s)}{}_{\a\b}^\r & 0 \\ \Qh^{(a)}{}_{\a\b}^\r & 2 \d_{\a\b}^{\r\s} \end{pmatrix} \begin{pmatrix} c_\r \\ \chi_{\r\s} \end{pmatrix} \\
   & = \int e \Big\{ \cb^\m (\na^2 \d^\r_\m +R^\r_\m + \f{1-\r}{2}\na_\m \na^\r) c_\r + 2 \, \bar\chi^{\m\n}\chi_{\m\n}  \\
   & \qquad \quad + \bar\chi^{\m\n} (\na_\m c_\n - \na_\n c_\m + 2\, c^\r \om_\r^{IJ} e_{\m I} e_{\n J} ) \Big\} \, .
\end{split}
\ee
The $\bar\chi C$ cross-term can be dropped as it does not contribute to the Faddeev-Popov determinant, thanks to the triangular structure of the matrix.

\subsection{Quadratic variation}

We now come to the expansion of the action. For simplicity, we disregard from now on the topological terms, and fix $\Lambda=0$. This is sufficient for our present goals. 
Hence, we work from now on with the basic action
\be\label{S2}
S[e,\om] = -\f{1}{\k^2} \int \tr \left\{\left(\star+\f{\Id}{\g}\right) \, e\w e\w F[\om] \right\},
\ee
and the gauge-fixing and ghost terms \Ref{Sgf} and \Ref{Sgh}. 

We take the change of variables \Ref{pertdef} and expand to second order in the quantum fields. It is convenient to work with the following projections,
$$f_{\m\n}=f_\mu^I e_{\nu I}, \qquad w^{I,JK}\equiv e^{I\m} w_\m^{JK}.$$
Then, the second variation of the action can be written as
\be\label{H1}
S^{(2)} = -\f12 \int e \left\{ f_{\m\n} M_{11}{}^{\m\n \a\b} f_{\a\b} + 2 w^{I,JK} M_{12}{}^{\m\n}_{ I,JK} f_{\m\n} + w^{I,KL} M_{22}{}_{I,JK}^{A,BC} w_{A,BC} \right\},
\ee
where the kinetic operator $M_{ij}$ is a function of the background fields $e_\mu^I$ and $K^{I,JK}$, and only the off-diagonal term $M_{12}$ contains (first) derivatives.
To write $M_{ij}$ explicitly, we define the matrix 
\be
P_{A,BC}^{I,JK} \equiv \d^I_A \left( \Id + \star/\g \right)_{BC}^{JK}\, .
\ee
 We then have\footnote{Here and in the following, we will be rather blas\'e about raising and lowering the Euclidean indices $I,J$, etc.}
\bea \label{M11}
 M_{11}{}^{\m\n \a\b} &=& 
\f1{2e}  e^{\n}_I e^{\b}_J e^{\g}_C e^{\d}_L \eps^{\m\a\r\s} \eps^{AB}_{KL} P^{K,IJ}_{C,AB} 
\left( R_{\r\s\g\d} + 2\nabla_{\r} K_{\s \g\d} + 2 K_{\r\g\l} K_{\s}{}^\l{}_\d \right)
\\ M_{12}{}^{\a\b}_{I,JK} &=& e^\a_A e^\b_B \eps^{AQRS} \eps^{BQMN} P^{I,[JL}_{R,MN} 
\left( \d^{K]}_L e^\s_S \nabla_\s -2 K_{S,L}{}^{K]}\right) \label{M12}
\\ M_{22}{}_{A,BC}^{I,JK} &=& 4 P_{[B,C]A}^{[K,J]I} \label{M22}
\eea
with $\na_\m$ the covariant derivative associated to the Christoffel connection 
$\G^\l_{\mu\nu}(e)$ of the background metric.

In order to diagonalize the operator $M_{ij}$, and obtain a second-order operator suitable for the heat-kernel expansion, we make a field redefinition with trivial Jacobian, by leaving $f$ unchanged and defining
\be
w \to \wt = w + [M_{22}]^{-1} [M_{12} f].
\ee
This gives
\be \label{S22}
S^{(2)} = -\f12 \int e \left\{ f_{\m\n} ([M_{11}]- [M_{12}]^T [M_{22}]^{-1} [M_{12}]){}^{\m\n\a\b} f_{\a\b}  
+ \wt^{I,KL} M_{22}{}_{I,JK}^{A,BC} \wt_{A,BC} \right\},
\ee
where the transpose $[M_{12}]^T$ is the same as $[M_{12}]$ but with an opposite sign for the derivative.

Notice that $M_{22}$ is not diagonal in the algebraic indices. 
To find its inverse, we consider the irreducible components of the connection. Recall that $w_{A,BC}$ transforms as a tensor in the $so(4)\cong so(3)\oplus so(3)$ representation 
\be\label{omreps}
\bf{(1/2,1/2)\otimes[(1,0)\oplus(0,1)]
= (3/2,1/2)\oplus(1/2,3/2)\oplus(1/2,1/2)\oplus(1/2,1/2)}.
\ee
The decomposition into (parity-even) irreducibles is realized by the three orthogonal projectors
\be\label{Preps}
\bar{P}_{A,BC}^{I,JK} =  \d_A^I \d_{BC}^{JK} - \check{P}_{A,BC}^{I,JK}  - \hat{P}_{A,BC}^{I,JK},
\quad \check{P}_{A,BC}^{I,JK} =  \f23 \d_{A[C} \d_{B]}^{[J} \d^{K]I},
\quad \hat{P}_{A,BC}^{I,JK} =  \f16 \eps_{ABCD}\eps^{IJKD}.
\ee
Using these projectors and their symmetry properties, one can show that
\bea
w^{A,BC} M_{22}{}_{A,BC}^{I,JK} w_{I,JK} 
&=& w^{A,BC}  \left[ \Big(\Pb - 2  \Ps- 2\Ph \Big) P \right]_{A,BC}^{I,JK} w_{I,JK}.\label{M22bd}
\eea
In this block-diagonal form,  $M_{22}$ can be immediately inverted, to give
\be\label{M22inv}
M_{22}^{-1} = \Big(\Pb - \f12  \Ps- \f12\Ph \Big) P^{-1}.
\ee
As anticipated, singularities are present at $\g^2=1$, since
\be
(P^{-1})_{A,BC}^{I,JK} = \f{\g^2}{\g^2-1} \d^I_A \left( \Id - \f1{\g} \star \right)_{BC}^{JK}.
\ee

Collecting \Ref{M12} and \Ref{M22inv}, we have
\bea\label{panino}
[M_{12}{}^T M_{22}^{-1} M_{12}]^{\a\b\m\n} &=& e_A^\a e^\b_B e^\m_M e^\n_N \eps^{AEFT} \eps^{BEGH} \eps^{MQRS} \eps^{NQUV}
\\ \nn && \ \times P^{I,JL}_{F,GH} P^{-1}{}_{I,JK}^{P,CD} 
\Big(\Pb - \f12  \Ps- \f12\Ph \Big)_{P,CD}^{X,YZ} P_{R,UV}^{X,YW}
\\ \nn &&  \times 
\Big[-\d_{KL} e^\tau_T \nabla_\tau - 2 K_{T,LK}\Big] 
\Big[\d_{WZ} e^\s_S  \nabla_\s - 2 K_{S,WZ}\Big] 
\eea
The explicit evaluation of this expression is a rather tedious operation, which involves
contracting all the indices among the various projectors appearing in the second line.
The operation can be handled very efficiently using the algebraic manipulator Cadabra \cite{Peeters}.
We omit the extended result, which is very long and not directly relevant to our scopes, and report below only the parts of interest. 

The tensor algebra simplifies greatly for the terms with two covariant derivatives, since the first $P$ matrix multiplies its inverse. In particular, for the part symmetric in $\a\b$ and $\m\n$ we obtain
\be\label{DD}
\begin{split}
[M_{12}{}^T M_{22}^{-1} M_{12}]^{(\a\b)(\m\n)}_{2-\rm derivatives} \equiv & -2 ( g^{\m\a} g^{\b\n} - g^{\m\n} g^{\a\b} ) \nabla^2 - 2 \big( g^{\a\b} \nabla^\m \nabla^\n +g^{\m\n} \nabla^\a \nabla^\b   \big) \\
  &+ 4 g^{\a \m} \nabla^{\n} \nabla^\b + 2 g^{\m\a} R^{\b\n} - 2 R^{\a\m\b\n} 
  + \f{1}{\g} \eps^{\a\m\r\s} R^{\b\n}_{\r\s}, 
\end{split}
\ee
with symmetrization on $\a\b$ and $\m\n$ on the RHS implicitly understood. The Riemann tensors appear from commuting covariant derivatives.
Expression \Ref{DD}, together with the Riemann part of \eqref{M11}, gives the usual linearized Einstein tensor on an arbitrary background.\footnote{Modulo a term proportional to the equations of motion, coming from the use of tetrads -- rather than metric -- fluctuations. See the comment at the end of Sec.~\ref{Sec:HKexp}.}
In particular, the Riemann terms proportional to $1/\g$ cancel between the two expressions.

\subsection{Choosing the gauge}

For $\g^2\neq 1$, the integral over $w$ does not require any gauge-fixing. Let us then focus on the part of \Ref{S22} depending only on the the tetrad fluctuation $f$, and denote $\hat H_f$ its Hessian, including the gauge-fixing terms. 
Writing $f_{\m\n}=(1/2)s_{\m\n}+(1/2)a_{\m\n}$, we have
\be
f_{\m\n} \hat H_f^{\m\n\a\b} f_{\a\b} = 
\f14 s_{\m\n} \hat H_f^{(\m\n)(\a\b)} s_{\a\b}  + \f12 s_{\m\n} \hat H_f^{(\m\n)[\a\b]} a_{\a\b}
+ \f14 a_{\m\n} \hat H_f^{[\m\n][\a\b]} a_{\a\b}.
\ee
Without loss of generality, we can take the simplest gauge $\b=0$, and freeze completely the $a_{\m\n}$ field.
This limit can be smoothly obtained either by rescaling the $a_{\m\n}$ field, leading to its disappearence from the rest of the action in the limit, or by rewriting the gauge fixing action by means of an auxiliary field, which in the limit $\b\to 0$ becomes a Lagrange multiplier, implementing the condition $a_{\m\n}=0$.
Accordingly, we omit in the following the contributions with one or more $a_{\m\n}$ fields, 
and look only at the pure $s_{\m\n}$-part of the Hessian. 

The remaining Hessian for $s_{\mu\nu}$ can be decomposed according to its number of covariant derivatives, as
\be\label{Hess2}
\f14 \hat H_f = \f14( -M_{11} + M_{gf} + M_{12}{}^T M_{22}^{-1} M_{12} ) =
H_2(\na_\m) +H_1(\na_\m) +H_0,
\ee
where the two-derivative operator comes from \Ref{DD} as well as the gauge-fixing term \Ref{Sgf},
$H_1(\na_\m)$ comes from the mixed terms of \Ref{panino} with one contorsion and one covariant derivative,
and finally $H_0$ contains $M_{11}$, the terms of \Ref{panino} with two contorsions, and 
the non-derivative terms in \Ref{DD}.

Once the explicit form of the Hessian is known, one typically looks for a gauge in which
the wave solutions on a flat background are null, that is a gauge in which the
second derivatives appear only in the form $\na^2$. A differential operator satisfying this condition is called minimal.
Explicitly, our second-order operator is
\bea\label{H2}
H^{\a\b\m\n}_2 &= & 
-\f12 \left[ g^{\m(\a} g^{\b)\n} + \left( \f{(1+\r)^2}{8\a} -1\right) g^{\m\n} g^{\a\b} \right] \nabla^2 \\ \nn &&
+ \f14 \left(\f{1+\r}{\a} -2\right) \big( g^{\a\b} \nabla^\m \nabla^\n +g^{\m\n} \nabla^\a \nabla^\b   \big) 
+ \left(1-\f1\a\right) g^{\a (\m} \nabla^{\n)} \nabla^\b. 
\eea
The gauge choice that leads to a minimal second-order operator is the De Donder gauge
$\a = \r = 1$.
In such a gauge,
\be
H^{\a\b\m\n}_2 = - C^{\a\b\m\n} \na^2, \qquad
C^{\a\b\m\n} = 2 g^{\m(\a} g^{\b)\n} - g^{\m\n} g^{\a\b},
\ee
and the Hessian \Ref{Hess2} can be casted in the following form,
\be
\f14 \hat H_f = C\big[-\Id \na^2+B^\m \na_\m + X\big] \equiv C \tilde{H},
\ee
where $B^\m\na_\m$ and $X$ are just the previous tensors $H_1$ and $H_0$ contracted with the inverse of $C$.

\section{Heat-Kernel expansion}
\label{Sec:HKexp}

After the gauge-fixing, the 1-loop effective action takes the form
\bea\label{S4}
e^{-\G} &=& e^{-S_{\rm tot}} \int {\cal D}s \, e^{-\f12 \la s, C \tilde H s \ra }
\int {\cal D}\cb {\cal D} c\, e^{ -\f12 \langle \cb, H_{gh} c \rangle  }
\int {\cal D}w \, e^{ \f12 \langle w, M_{22} w \rangle }.
\eea
Since $[M_{22}]$ contains neither derivatives nor any dynamical field, the $w$ functional integral can be performed\footnote{Note that $-M_{22}$ is not positive definite for  any value of $\g$:
its eigenvalues are $-1\pm 1/{2\g}$, each with multeplicity eight, and
$4/3(5\pm {\sqrt{16 +9/\g^2)}})$, each with multeplicity four.
Therefore a proper definition of the integral will require either an analytical continuation of some of the components of $w$, or the addition of some other appropriate torsion-squared term, as suggested in \cite{Tseytlin}.} with the standard Gaussian measure, giving a trivial factor $1/\sqrt{\det M_{22}}$.
Doing so, one obtains a formal equivalence between the first and second order formalisms (at least as long as one considers only observables built from the tetrad and not the connection), along lines already appeared in the literature (e.g. \cite{Tseytlin,FTdual,Aros}).\footnote{The equivalence clearly relies on the invertibility of the tetrad. As mentioned earlier, we do not look here at possible effects arising from contributions of degenerate tetrads, which are expected to spoil the equivalence. }

Concerning the integral over the metrics, the standard De Witt's prescription (see also \cite{BernBlauMottola})
is to normalize with respect to the supermetric $C$, 
\be\label{DWmeasure}
\int Ds_{\m\n}\, e^{-\f12  \, \langle s, C s\rangle } = 1.
\ee
If we do so, we read from \Ref{S4} the 1-loop effective action
\be
\Gamma_{\rm 1-loop} = \f12 \Tr \ln \tilde H - \Tr \ln H_{gh}. 
\ee
Both operators are of the minimal form $-\na^2 + B^\m \na_\m + X$,
and the heat-kernel expansion for such operators is given for example in \cite{Barvinsky:1981rw, Barth:1985cz}. The first coefficient $a_0$ is just the identity $\Id$ over the internal indices, and the second coefficient reads
\be
a_1 = \f16 R \,\Id  + \f12 \nabla_\m B^\m - \f14 B_\m B^\m - X,
\ee
where $R$ is the Ricci scalar.

For the ghost contribution, the internal space is the vector space, and the Hessian can be read from \Ref{Sgh}. In the chosen gauge, it reduces to 
\be
H_{gh}^{\m\n} = - g^{\m\n}\na^2 - R^{\m\n}.
\ee
Therefore,
\be
\tr \, a_0{}_{gh} = 4\,  , \qquad
\tr \, a_1{}_{gh} = \f{5}{3} R \, .
\ee

For the gravitational contribution, the internal space are the symmetric indices $(\m\n)$.
Disregarding total derivatives, the non-trivial part of the calculation amounts to computing the traces of the operators 
$B_\m B^\m$ and $X$. This we did starting from \Ref{panino} and using Cadabra to contract the indices. We found the following expressions,
\bea
\tr[X] &=& 4 R +\f32 K_{\m\n\r}K^{\m\n\r} + \f32 K^\m{}_{\m\r}K_{\n}{}^{\r\n} + \f32 K_{\m\n\r}K^{\n\m\r} \\\nn
      && \quad +\f{1}{e} (\f{9}{2\g} \eps^{\m\n\r\s} K_{\m\n}{}^{\l} K_{\r\s\l} +\f{3}{\g} \eps^{\m\n\r\s} K_{\m\n\r} K^\l{}_{\l\s}), 
\\
\tr[B^\m B_\m] &=& \f{5-3\g^2}{\g^2} K_{\m\n\r}K^{\m\n\r} + \f{3-13\g^2}{\g^2} K^\m{}_{\m\r}K_{\n}{}^{\r\n} + \f{3\g^2-13}{\g^2} K_{\m\n\r}K^{\n\m\r} \\\nn
      && \quad +\f{1}{e} (- \f{6}{\g} \eps^{\m\n\r\s} K_{\m\n}{}^{\l} K_{\r\s\l} +\f{4}{\g} \eps^{\m\n\r\s} K_{\m\n\r} K^\l{}_{\l\s}).  
\eea

Putting these results together, the divergent part of the 1-loop effective action, equation \Ref{Gammadiv}, gives
\be\label{EA}
\Gamma^{\rm div}_{1-\rm loop} = -\frac{1}{32 \pi^2} \bigg\{ \L_{UV}^4 \int e
 - {\L_{UV}^2} \int e {\cal L}_1  - \ln(\L_{UV}^2/\m^2) \int e  {\cal L}_2 \bigg\},
\ee
with
\bea \label{L1}
{\cal L}_1 &=& \f{17}{3} R(e) + \f{3\g^2+5}{4\g^2} K_{\m\n\r}K^{\m\n\r} 
 - \f{7\g^2-3}{4\g^2} K^\m{}_{\m\r}K_{\n}{}^{\n\r} + \f{9\g^2-13}{4\g^2} K_{\m\n\r}K^{\n\m\r} \nn\\
 && +
 \f{3}{\g e} \eps^{\m\n\r\s} K_{\m\n}{}^{\l} K_{\r\s\l} +\f{4}{\g e} \eps^{\m\n\r\s} K_{\m\n\r} K^\l{}_{\l\s}.
\eea
As expected, all the dimension-two invariants have appeared in the quadratic divergences (cf. \Ref{dim2}).
${\cal L}_2$ contains the dimension-four operators coming from the coefficient $a_2$. This is a very long expression which will not be needed in the following.

The result can be compared with the similar calculation performed in the metric second-order formalism. 
The relation of the tetrad perturbation to the usual metric perturbation is given by
\be \label{h}
h_{\m\n} = s_{\m\n} +f_\m^I f_{\n I},
\ee
according to $g_{\m\n}\to g_{\m\n}+h_{\m\n}$ and $g_{\m\n} = e_\m^I e_\n^J \d_{IJ}$.
The quadratic piece in \Ref{h} contributes to the operator $H_0$ in \Ref{Hess2} an extra factor 
$-2 g_{\a(\m} G_{\n)\b}$, where $G_{\m\n}=R_{\m\n}-\f12 g_{\m\n}R$.
This contributes an extra $\Tr[-2 (C^{-1})_{\m\n}^{\r\s} g_{\a(\r} G_{\s)\b}]=2R$ to $\tr[X]$.
Adding this contribution to \Ref{EA} and setting $K=0$, the result agrees with what we find in the second-order formalism \cite{Fradkin:1978yf}. 
This provides a consistency check of our calculations.
%

\section{Renormalization}
\label{Sec:Renorm}

We have isolated the contributions to the divergent part of the effective action. In this Section, we discuss how to renormalize it.
It is well known in quantum field theory that the $S$-matrix is unaffected by local field redefinitions in the effective action. As a consequence, any term in the effective action which vanishes on-shell does not contribute to the $S$-matrix. This is because we can always write such a term as proportional to the equations of motion, and then we can reinterpret it as coming from an infinitesimal field redefinition. In this sense, divergences which vanish on-shell are innocuous, as they can be eliminated by field redefinitions. Furthermore, these same divergences are also gauge-dependent \cite{BuchbinderOdintsovShapiro}, and gauges can be found where they are absent \cite{Kallosh}.

Pure quantum gravity at 1-loop turns out to be on-shell finite in all its formulations \cite{'tHooft,ChristensenDuff,BuchbinderShapiro}, and the present one is no exception.
Indeed, having assumed invertibility of the tetrad, the equations of motion for \eqref{S2} reduce to
\be
R_{\m\n}=0, \qquad K_{\m\n\r}=0,
\ee
and all the quadratic and logarithmic\footnote{Even if we did not exhibit the $a_2$ coefficient, notice that any non-topological term in it must be proportional to either $R_{\m\n}$ or $K_{\m\n\r}$. The only exception would be $R_{\m\n\r\s}R^{\m\n\r\s}$, but this can be eliminated by using the Gauss-Bonnet invariant \cite{'tHooft}.} divergences in \eqref{EA} vanish on-shell.
Only the quartic divergence survives, but there are at least two ways to deal with it. A natural one is to keep, as in \cite{ChristensenDuff}, $\L\neq 0$ in the bare Lagrangian and reabsorb the quartic divergence in the renormalized $\L$. The second, is to modify the measure \Ref{DWmeasure}, 
carefully taking into account $\d^{(4)}(0)$ terms, as suggested by Fradkin and Vilkovisky, with the consequence that the quartic divergences are automatically cancelled \cite{FV}. 

As well-known, the situation changes at higher order. For spacetime dimensions greater than three, $R_{\m\n}=0$ does not fix all the components of the Riemann tensor, hence one finds that at 2-loops the $R_{\m\n}{}^{\r\s}R_{\r\s}{}^{\a\b}R_{\a\b}{}^{\m\n}$ divergence survives on-shell, thus establishing the non-renormalizability of gravity \cite{GoroffSagnotti}.
While such terms will certainly be present also in the first-order formulation we are considering here, it seems not possible to have new non-renormalizable terms, as the equation $K_{\m\n\r}=0$ fixes all the components of the contorsion.\footnote{Since we are including parity-odd terms, one may expect also cubic combinations of the Riemann tensor with one or more epsilon tensor to give rise to new non-renormalizable terms. However, these either vanish on-shell or are proportional to topological terms \cite{Kallosh74}.} Thus, the equivalence between the two formulations seems to persist at the quantum level despite the non-renormalizability of the theory.

\subsection{Off-shell running of the coupling constants}
\label{Sec:Offshell}

The situation will generally change in presence of a source for curvature or torsion, as we will discuss in the following section.
As a first step in that direction, one might consider a different renormalization scheme for the pure gravity case, in which some of the divergences are reabsorbed into a redefinition of the couplings. Specifically, we can consider a scheme in which we reabsorb into coupling renormalizations all the divergences for which this can be done, while for the others we use a field redefinition or introduce appropriate counterterms. Such a scheme has been used e.g. in \cite{Fradkin:1978yf,BuchbinderShapiro}, and it also comes closer to the spirit of the calculations done in the context of the asymptotic safety scenario, where the running of the traditionally inessential Newton's constant can be motivated by the special role it has in the theory \cite{PercacciPerini}.
With these considerations in mind, we now concentrate on the quadratic divergences and proceed to examine the structure and consequences of such a renormalization scheme.

The quadratic divergences proportional to the Ricci scalar (first term of the first line of \Ref{L1}) and to the Holst term (first term of second line of \Ref{L1}) can be absorbed by a non-minimal subtraction ansatz (see for example \cite{Robinson, Niedermaier}):
\begin{subequations}\label{nonmin}\bea
\f{1}{\k^2_R} &=& \f{1}{\k^2} \left( 1 -  \f{17}{3}\frac{1}{32 \pi^2} \k^2 \L_{UV}^2 + b_1 \right),
\\
\f{1}{\g_R \k^2_R} &=& \f{1}{\g \k^2} \left(1 -  \frac{3}{32 \pi^2} \k^2 \L_{UV}^2 + b_2 \right).
\eea\end{subequations}
The requirement of cancellation of the divergences leaves the finite coefficients $b_1$, $b_2$ unconstrained.
The usual minimal ansatz corresponds to taking $b_1=b_2=0$, and it leads to renormalized couplings with no dependence on any renormalization scale $\m$.
This is the choice made by several authors in the literature, and of course it means that quadratic divergences lead to no running of the couplings \cite{Anber}.
Alternatively, one can introduce a renormalization scale (or sliding scale \cite{WeinbergV2}) $\m$ by using a different prescription.
This can be done in a consistent way with what one is forced to do at higher orders of the heat kernel expansion. There, we meet also IR divergences, and these can be regulated by cutting-off the upper limit of the $t$-integration at $1/\m^2$, for $\m < \L_{UV}$, as we have already implicitly done when writing the logarithmic divergence as $\ln(\L_{UV}^2/\m^2)$. Keeping $\m>0$ also for the first two (IR-finite) terms of the heat kernel expansion, we find finite $\m$-dependent terms that can be naturally absorbed in the redefinition of the couplings. In our case, this procedure leads to \Ref{nonmin} with
\be
b_1 = \f{17}{3}\frac{1}{32 \pi^2} \k^2 \m^2, \qquad b_2 = \frac{3}{32 \pi^2} \k^2 \m^2.
\ee
As emphasized in \cite{Niedermaier}, this type of ansatz is very natural from the point of view of the Wilsonian renormalization group, as it corresponds to implementing the matching conditions
\be
\k^2_R(\L_{UV}=\m) = \k^2, \qquad  \g_R(\L_{UV}=\m) = \g,
\ee
stating the fact that the bare couplings represent the initial condition for the flows.

The non-minimal subtraction ansatz allows us to define non-trivial beta functions from quadratic divergences. The beta functions are written for the dimensionless coupling constants, hence we define the dimensionless Newton's constant $g\equiv\f{1}{16\pi}\m^2\k^2$. 
After rewriting the renormalization conditions as 
\be
g_R = g (1 +  \f{17}{6 \pi} g \f{\L_{UV}^2}{\m^2} - \f{17}{6 \pi} g ) \, ,
\ee
\be
\f{1}{\g_R} = \f{1}{\g } (1 +  \frac{4}{3 \pi} g \f{\L_{UV}^2}{\m^2} -\frac{4}{3 \pi} g ) \, ,
\ee
we can finally obtain the beta functions
\begin{align}
& \b_g(g_R) = \m\f{\p g_R}{\p \m} = g_R ( 2 -  \f{17}{3 \pi} g_R ), \\
& \b_\g(g_R) = \m\f{\p \g_R}{\p \m} = \frac{4}{3 \pi} \g_R g_R.
\end{align}

Some observations can now be made. 
First, we note that the beta function for Newton's constant is independent of $\g$.\footnote{A $\g$-dependent renormalization of $G$ has been considered in \cite{Jacobson}. Notice that one could decide to reabsorb into $G$ the divergences of one of the last two terms in the first line of \Ref{L1} (see \eqref{epseeF}), instead of that proportional to the Ricci scalar as we did. 
This would lead to a $\g$-dependent running of $G$. It would be interesting to see whether this could have any bearing on the argument of \cite{Jacobson}.}
Second, $\g=0$ and $\g=\infty$ are fixed points, consistently with the result claimed in \cite{DaumReuter}. We find that the points are respectively IR- and UV-attractive. If this were a physical feature, it would be nicely consistent with the idea that the metric correctly captures the degrees of freedom of general relativity at low energies, while the connection field becomes more important at high energies.
On the other hand, the points $\g=\pm1$ seem to have no role.
Finally, we also find a non-Gaussian fixed point for the Newton constant, in agreement with the asymptotic safety conjecture \cite{Weinberg79,NiedermaierReuter,Codello:2008vh,Litim,BMS}, but of course this goes beyond the realm of perturbation theory.

As already stressed, the running we just presented for $\g$ is scheme and gauge dependent, hence not physical. We will need to add a source of torsion in order to render it physical and test whether any of the qualitative features we found persists. This is what we do in the next Section.

\section{Adding fermions}
\label{Sec:Fermi}

The situation becomes more interesting if one includes torsion-generating matter, such as fermions.
In this case, we expect the theory to be non-renormalizable even at one loop \cite{DeserNieuwenhuizen,Barvinsky:1981rw},
and to get a physical on-shell running for the Immirzi parameter.

\subsection{Fermion action and effective interaction}

At the classical level, the effect of fermions coupled to the action \eqref{S2} has been discussed to a great extent, see in particular \cite{Perez:2005pm,Freidel:2005sn,Mercuri:2006um,Bojo,Alexandrov:2008iy,Cianfrani:2010zm}.
The key point is that using \Ref{omdec}, the field equations simply fix the on-shell value of the contorsion $K$ to be a unique combination of fermion currents. Therefore, the connection can be effectively integrated out, and the coupled system reduces to second-order tetrad gravity (with no torsion) coupled to fermions, with the addition of an interaction term coming from the on-shell value of $K$. Let us briefly review the relevant details.
We consider the following action \cite{Freidel:2005sn},\footnote{For consistency with the rest of the paper we keep here the Euclidean signature. For the definition of fermions in Euclidean space we follow the construction of \cite{Mehta:1986mi,vanNieuwenhuizen:1996tv}. Our conventions are
\be
\{ \g^I,\g^J\}=2 \d^{IJ}, \qquad (\g^I)^{\dagger}=\g^I, \qquad (\g^I)^2=1 \, ,
\ee
\be
\g^5=i \g_1\g_2\g_3\g_4, \qquad \{ \g^5,\g^I\}=0, \qquad (\g^5)^{\dagger}=-\g^5, \qquad (\g^5)^2=-1 \, .
\ee
We have $\psib=\psi^{\dagger}\g^5$, which ensures $so(4)$-invariance of the action.
Other useful formulas are
\be
\{ \g^K, [\g^I,\g^J]\} = -i 4 \eps^{IJKL} \g^5 \g_L,
\qquad [\g^K,[\g^I,\g^J]] = 8 \d^{K[I} \g^{J]}.
\ee
}
\be \label{Sfermi}
S_{\psi}[e,\om,\psi,\psib] = - \f{i}{4} \int d^4 x\, e \left( (1-i\th) \psib \g^I e_I^\m \na_\m \psi  - (1+i\th) \overline{\na_\m \psi} \g^I e_I^\m\psi  \right),
\ee
where
\be
\na_\m \psi = \p_\m\psi +\f18 \om_{\m IJ}[\g^I,\g^J] \psi.
\ee
For $\th=0$ we recover the standard minimal coupling. The interest of the non-minimal coupling will become clear below.
Notice that if the connection $\om_{\m}^{IJ}$ was torsionless, $\psib \g^I e_I^\m \na_\m \psi  + \overline{\na_\m \psi} \g^I e_I^\m\psi$ would be a total derivative,
and as a consequence $i\psib \g^I e_I^\m \na_\m \psi $ would be Hermitian. Because of the presence of torsion, this is not true anymore, hence one needs to keep both terms in \Ref{Sfermi}, and the non-minimal term proportional to $\th$, becomes non-trivial.

If we use the decomposition \eqref{omdec} in the coupled gravity-fermion system, the total action \Ref{S2} plus \Ref{Sfermi} is quadratic in $K$, and so we can immediately solve the field equations for it. 
A simple calculation gives
\bea\nn
\f1e \f{\d S_\psi}{\d \om_{\m}^{IJ} } &=& - \f12 e^{\m K} \left( \f14 \eps_{IJKL} A^L +\f{\th}{2} \d_{K[I} V_{J]}   \right),
\\\nn
\f1e \f{\d S}{\d \om_{\m}^{IJ} } &=& 
- \f{1}{\k^2} e^{\m K} \left[M_{22} \right]^{A,BC}_{K,IJ} K_{A,BC},
\eea
from which we find, for $\g^2\neq 1$, the following unique solution,
\be\label{Kbar}
\begin{split}
\bar K_{I,JK}(\psi) 
 &= \f{\k^2}8 \f{\g^2}{\g^2-1} \left( \f12 \eps_{IJKL} (A^L-\f{\th}{\g}V^L) - \f{1}{\g} \d_{I[J} (A_{K]} - \th \g V_{K]} )  \right),
\end{split}
\ee
expressed in terms of the fermionic vector and axial currents,
\be
V^I = \psib \g^I \psi, \qquad A^I=\psib \g^I \g^5 \psi.
\ee
Hence, the reduced action takes the form
\be \label{Scoupled}
S_{\text{coupled}}[e,\psi]  = S[e]+S_\psi [ \psi,e] + S_{\text{int}}[\psi, e],
\ee
where both $S$ and $S_\psi$ are evaluated at $\om=\om(e)$, which in particular reduces $S$ to the standard Einstein-Hilbert action
\be \label{S-EH}
S[e] = -\f{1}{\k^2} \int d^4 x \, e R(e),
\ee
and $S_\psi$ to the standard Dirac action
\be \label{Sfermi2}
S_\psi [ \psi, e] = - \f{i}{2}  \int d^4 x\, e \, \psib \g^I e_I^\m \na_\m(e) \psi.
\ee
The extra interaction term is 
\be\label{Sint}
\begin{split}
S_{\text{int}} & = \f{1}{2\k^2} \int  d^4 x\, e \bar{K}^{K,IJ} \left[M_{22} \right]^{A,BC}_{K,IJ} \bar{K}_{A,BC} \\
  & = - \f{3\k^2}{128} \Big( \f{\g^2}{\g^2-1} \Big) \int  d^4 x\, e \Big(  A_I A^I - 2\f{\th}{\g} A_I V^I + \th^2 V_I V^I    \Big).
\end{split}
\ee
This is where all the dependence on the Immirzi parameter goes.
The bottom line is that taking the connection as an independent variable amounts to simply adding the four-fermion interaction \Ref{Sint} to the usual second order formalism. 
The Immirzi parameter has now acquired a physical meaning, as it enters explicitly the coupling between the gravitational field and the fermionic currents. 
Its ``interpolating'' role mentioned in Section \ref{Sec:Action} is also clearer:
for $\g=0$ the solution to the equations of motion is $\bar{K}_{I,JK} = 0$, and there is no four-fermion interaction, while for $\g=\infty$ we find instead what expected in the Einstein-Cartan case.

An interesting feature of $S_{\rm int}$ is the presence of the parity-odd term $A_I V^I$.
This term is responsible for gravity-induced parity breaking in the fermionic sector, but it is hugely unconstrained by current observations \cite{Freidel:2005sn}, because of the weakness of the gravitational coupling as well as the large effect already caused by the weak interactions. Notice that there is no parity breaking effect for the minimal coupling $\th=0$. This might look at first puzzling, since the initial Holst term was parity odd. 
However, the assessment that the Holst term is parity odd is made under the assumption of a parity-even torsion. On the contrary, we see from \Ref{Kbar} that the on-shell contorsion $\bar K$ has an undefined parity. Then, the Holst term can a priori acquire both signatures on-shell, and as it turns out from the interaction term \Ref{Sint}, for $\th=0$ only a parity-even contribution remains.

For $\g^2=1$, we obtain gravity in self-dual variables \cite{Ashtekar,Plebanski}, and the only solution from varying the connection is $K=0$ and $\psi=0$. That is, fermions can not be consistently coupled to gravity in self-dual variables using \Ref{Sfermi}. An action to couple fermions to self-dual gravity is given in \cite{Romano}. In the following, we will not be interested in this case, and assume $\g^2\neq 1$ as we did in the previous sections.\footnote{On the other hand, it has also been remarked in \cite{Mercuri:2006um, Alexandrov:2008iy} that the physical effects of the Immirzi parameter due to its presence in \Ref{Sint}, are equivalent to taking a further non-minimal coupling of the fermions, with $(1-i\th)\Id$ in \Ref{Sfermi} replaced by $(1-i\th)\Id + (\tau-i\rho)\g^5$. With this starting point, the gravity-fermion system alone can not operationally distinguish between $\g$ and the other parameters $\th, \tau$ and $\rho$.
In particular in \cite{Mercuri:2006um} it has been suggested the special choice $\th=\t=0$, $\rho=1/\g$, which has the advantage that in the limit $\g^2\to 1$ it reproduces the action of \cite{Romano}. The flipside is that it corresponds to a fine-tuning of the contorsion such that the Holst term reduces to the Nieh-Yan invariant, thus dropping the Immirzi parameter out of the theory again. Furthermore, as we will show below this special coupling is unstable under radiative corrections. \label{footMercuri}}  

To complete this section on the classical theory, we give the remaining field equations.
From the variation of $\psib$ we get
\be
-i \g^I e_I^\m\na_\m\psi - \f{3\k^2}{32} \f{\g^2}{\g^2-1} \Big(\g^5 \g^I \psi A_I + \th^2 \g^I\psi V_I -\f{\th}{\g} \g^5\g^I\psi V_I -\f{\th}{\g} \g^I \psi A_I      \Big) = 0,
\ee
while the Einstein equations are
\be\label{EEf}
G_{\m\n} = -\f{\k^2}{2} T_{(\m\n)}, \qquad 0 = T_{[\m\n]},
\ee
with
\be
T^{\m\n} = \f1e \f{\d(S_\psi+S_{\text{int}} )}{\d e_\m^I} e^{I\n}.
\ee
The symmetric part of the (on-shell) energy-momentum tensor is given by
\be
T^{(\m\n)} =
 \f{i}{4} \Big( \psib \g^L e_L^{(\m} \na^{\n)} \psi -  \overline{\na^{(\n} \psi} \g^L e_L^{\m)} \psi   \Big) + \f{3\k^2}{128} \f{\g^2}{\g^2-1} g^{\m\n}  \Big(  A_I A^I - 2\f{\th}{\g} A_I V^I + \th^2 V_I V^I    \Big).
\ee
Notice that it has acquired a trace, although we are working with massless fermions. This is a consequence of the non-zero on-shell contorsion.

Moving on to the quantization of the coupled system, we could choose to start either with the action \eqref{S2}+\eqref{Sfermi}, or with the partially on-shell \eqref{Scoupled}.\footnote{The reader familiar with supergravity will recognize the latter as the so-called 1.5 formalism.}
As we are now interested in the on-shell renormalization, and since the contorsion field is non-dynamical, it turns out that the two choices lead to identical results.
For a matter of simplicity, we will choose to work with the latter form of the action.
Before going to the details, we need to discuss some delicate issues on the renormalization of coupling constants in general relativity.

\subsection{Essential and inessential couplings}
%
By definition, a coupling constant is \emph{inessential} if it can be removed by a field redefinition, and so it does not enter in the S-matrix.
Following \cite{Weinberg79}, we know that a coupling $g$ is {inessential} if and only if $\p S/\p g$ vanishes when we use the equations of motion.
Our Lagrangian \Ref{Scoupled} has the structure
\be
\LL = \eta \Big(  -\f{1}{\k^2} R -\f{i}{2} \psib \g^I e_I^\m \na_\m \psi -\f{\k^2}{2} (x A^2 + y V^2 + z AV ) \Big).
\ee
On-shell we find that $\p\LL/\p\eta = \p\LL/\p\k^2 = 0$, that is, both $\eta$ and $\k^2$ are inessential couplings. 
This means that we are allowed to set $\eta=\k^2=1$ in our Lagrangian, since they can be removed by trivial field redefinitions. To see this, we first rescale $\psi\to \f{1}{\k}\psi$, obtaining
\be
\LL = \f{\eta}{\k^2} \Big(  - R -\f{i}{2} \psib \g^I e_I^\m \na_\m \psi -\f{1}{2} (x A^2 + y V^2 + z AV )  \Big).
\ee
A further rescaling $e_\m^I\to \f{\k}{\sqrt{\eta}} e_\m^I$, $\psi \to \f{\eta^{1/4}}{\k^{1/2}}\psi$
 eliminates the overall coupling from the action.

On the contrary, $x$, $y$ and $z$ are all essential couplings, and can not be eliminated. Specifically,
\be \label{xyz}
x= \f{3}{64} \f{\g^2}{\g^2-1}, \qquad
y = \th^2 x, \qquad
z= -2\f{\th}{\g} x.
\ee
We see that we have only two variables, $\th$ and $\g$, for the three essential couplings. On the other hand, radiative corrections will generically produce all three terms $A^2$, $V^2$ and $AV$, which will lead to quadratic divergences. Hence, we will not be able to reabsorb the divergences into redefinitions of the sole two couplings $\g$ and $\th$. 
Assuming that some magic cancellations do not occur, we conclude that the quadratic divergences of the non-minimally coupled system are not renormalizable at one loop.

A possible solution is to start with the other $K^2$ terms in the bare gravitational action, and hope to get in this way three independent $x$, $y$ and $z$.\footnote{The same considerations apply to the most general non-minimal coupling of footnote \ref{footMercuri}: in this case, as shown in \cite{Alexandrov:2008iy}, the action really depends on just three parameters, but now we would have four essential couplings to renormalize, as there is an additional one in the kinetic term.}  Such terms are of course natural to add, since we are not insisting in the non-invertibility of the metric.
However, a simpler solution is to fix $\th=0$. In this way we are left with a single essential coupling $x$. In the following, we will consider only the minimally coupled theory, and show that the quadratic divergences can be renormalized using simply the Holst action.

\subsection{1-loop results}

For $\th=0$, the action is given by 
\be\label{Scoupled2}
S_{\text{coupled}}[e,\psi] = -\int d^4 x \, e 
\left\{\f{1}{\k^2}  R(e) + \f{i}{2} \psib \g^I e_I^\m \na_\m(e) \psi + \f{3\k^2}{128} \Big( \f{\g^2}{\g^2-1} \Big) A_I A^I \right\}.
\ee
This action coincides with the one considered by Barvinsky and Vilkovisky \cite{Barvinsky:1981rw}, with their coupling constant $\a$ now a function of the Immirzi parameter, $\a\equiv {\g^2}/({\g^2-1})$.
The 1-loop calculation for \eqref{Scoupled2} has already been performed in \cite{Barvinsky:1981rw}, and we can largely draw from their results. The calculations of \cite{Barvinsky:1981rw} were done for Lorentzian signature, but luckily, it is not too difficult to adapt their result to our Euclidean choice, by carefully following their steps and taking care of the different signs (in particular when $\eps$ tensors are being contracted). The final results, i.e. the beta functions, actually turn out to be independent of the signature.

The general 1-loop calculation is of the type outlined in Sec.~\ref{Sec:1loop}. The two new ingredients introduced in  \cite{Barvinsky:1981rw} were a fermion-dependent modification of the gauge-fixing, needed to maintain the gravitational Hessian in the form of a minimal operator, and a procedure to ``square'' the fermionic Hessian. These require the fermions to be Majorana spinors. We refer to the original paper for details, and assume from now on that we are dealing with Majorana spinors. We use their results for the 1-loop effective action in our set-up, and postpone the possible extension to Dirac spinors to future work. 

The 1-loop effective action has again the structure \Ref{EA}. After adapting the results in \cite{Barvinsky:1981rw} for the change of signature and the presence of the Immirzi parameter, the quadratically-divergent part reads
\be \label{1loop-offshell}
{\cal L}_1 = \f{11}{2} R + \f{3}{512} \left(6\f{\g^2}{\g^2-1} -5\right) \k^4 A^2,
\ee
where the equations of motion for the fermions have already been used. Using also the Einstein equations for the tetrad, we find
\be \label{1loop-onshell}
{\cal L}_1 = \f{3}{512} \left(28 \f{\g^2}{\g^2-1} -5\right) \k^4 A^2.
\ee
As for the logaritmic divergences, the final on-shell result is 
\be\label{L2}
{\cal L}_2 = \f12\left(3\f{\g^4}{(\g^2-1)^2} + 4\f{\g^2}{\g^2-1} +13\right) R_{\m\n} R^{\m\n}.
\ee

Unlike in the vacuum case, the logarithmic divergences are now of crucial importance: they do not vanish on-shell,
and can not be reabsorbed in a renormalization of the bare couplings. Therefore, the theory is non-renormalizable, as expected, and can only be made sense of as an effective field theory.\footnote{The logaritmic divergences cancel for  $\g^2=3/4\pm i \sqrt{7/5}$, but these are unphysical complex values, and furthermore they are not stable under renormalization, as we show below.}
Our focus here is not on the non-renormalizability, which was not in doubt, but rather the fate of the Immirzi parameter.
To explore any consequences for the role played by the Immirzi parameter in this theory, we look at the quadratic divergences \Ref{1loop-onshell}. 
With a similar non-minimal ansatz as in Sec.~\ref{Sec:Offshell}, but this time on-shell, these can be reabsorbed into the following renormalization of the Immirzi parameter,
\be\label{gR}
\f{\g_R^2}{\g_R^2-1} = \f{\g^2}{\g^2-1} - \f{1}{128\pi^2} (\L_{UV}^2-\m^2)\k^2 \left(28  \f{\g^2}{\g^2-1} -5 \right).
\ee

By inspection, we see that neither $\g=0$ nor $\g=\infty$ are stable under renormalization. 
It is of particular interest what happens if one starts with vanishing bare Immirzi parameter, $\g=0$. The initial action \Ref{Scoupled2} reduces to the second-order Einstein-Hilbert action coupled to fermions, with no four-fermion interaction. However, the latter is nonetheless generated by radiative corrections, see \Ref{1loop-onshell} with $\g=0$.
Namely, the radiative corrections introduce quadratic divergences which are  non-renormalizable in the second-order formalism.
In order to renormalize these divergences one is forced to introduce the four-fermion term in the classical action, that is, one is forced to have a non-vanishing Immirzi parameter.
In this sense, the first order formulation is more suitable to quantize the coupled gravity-fermion system.

From \Ref{gR}, we obtain the beta function of the Immirzi parameter,
\be \label{beta-onshell}
\m\f{\p \g^2_R}{\p \m} = - (\g^2_R-1) \f{\m^2 \k^2}{(8\pi)^2} (23\g^2_R+5).
\ee
The equation can be easily solved, leading to
\be \label{sol-onshell}
\g_R^2(\m) = \frac{ (23\g_0^2+5) e^{\frac{7g_0}{2\pi}(\m^2/\m_0^2-1)} +5 (\g_0^2-1) }{ (23\g_0^2+5) e^{\frac{7g_0}{2\pi}(\m^2/\m_0^2-1)} -23 (\g_0^2-1) },
\ee
where $g_0=G\m_0^2$, and $\g_0=\g_R(\m=\m_0)$ is the initial condition.
As the effective interaction in \eqref{Scoupled2} only depends on the effective coupling $\tfrac{\g^2}{\g^2-1}$,
we have to be careful if we want $\g_R(\m)$ to remain real for all values of $\m$. That is, we must take care that the right hand side of \eqref{sol-onshell} does not become negative.
It turns out that this requirement imposes some restriction on the initial conditions.
Defining $r=e^{\frac{7g_0}{2\pi}}$, we find that for $\g_0^2>1$ one has to take $\g_0^2<\tfrac{r+5/23}{r-1}$ so that the denominator of \eqref{sol-onshell} does not become negative for $\m\to 0$,
while for $\g_0^2<1$ one has to take $\g_0^2>\tfrac{r-1}{r+23/5}$ for the numerator not to become negative in the same limit.
Note that, given these bounds, in the IR limit $\m\to 0$ the Immirzi parameter flows towards a value between $\g_0$ and $|\g_R|=+\infty$, for $\g_0^2>1$, or between $\g_0$ and $\g_R=0$, for $\g_0^2<1$. The precise value depends on the initial condition. The fixed points $\g=0$ and $\g=\infty$ found in the pure gravity case are never reached.
On the contrary, we find that for any initial condition the point $\g_R^2=1$ is reached in the UV limit $\m\to\infty$.

Three remarks can be made from the expressions of the beta function and its solution. First, we have two independent sectors, for $|\g_R|$ larger or smaller than 1, with same UV limit and opposite IR limit.
Second, the only real fixed points are at $\g^2=1$, but before attributing much significance to such fixed points, one should bear in mind that they correspond to a divergent coupling for the four-fermion interaction, hence they are out of the range of validity of perturbation theory.
Third, we have an explicit dependence of the beta function on the external parameter $\m^2 \k^2$. This can be interpreted as the renormalization scale measured in Planck units, as we are not letting $\k^2$ run in the present scheme. The explicit appearence of the renormalization scale in the beta function is a manifestation of what discussed in  \cite{PercacciPerini}, that when treating Newton's constant as an inessential parameter, we will usually obtain non-autonomous systems of renormalization equations
(Notice that this phenomenon does not take place at the level of logaritmic divergences, and thus it does not appear in dimensional regularization).

Alternatively, we could again follow \cite{PercacciPerini}, as we did above in Sec.~\ref{Sec:Offshell}, by renormalizing $\k^2$ and studying its running. A running for Newton's constant can be introduced with a partially off-shell scheme, using \eqref{1loop-offshell} rather than \eqref{1loop-onshell} to renormalize both $\k^2$ and $\g$. Such a procedure leads to an autonomous system, but it does not change the qualitative conclusions about the fixed points for $\g$, which are on-shell results. Indeed, the running of Newton's constant turns out to be again independent of $\g$, with beta function $\b_g = g_R (2- \tfrac{11}{2\pi}g_R)$.
Its only effect on \eqref{beta-onshell} is to replace $\tfrac{\k^2\m^2}{16\pi}= g_0 \m^2/\m_0^2$ with a non-trivial running $g_R(\m)$, bounded between $g_R(\m=0)=0$ and $g_R(\m=\infty)=\tfrac{4\pi}{11}$. This modifies things like the velocity along the flow and the reality bounds on $\g_0$, but not the conclusions about the special points $\g_R^2=0,1,\infty$.

To conclude, we would like to stress that the system we considered here corresponds to the simplest coupling between the Holst action and fermions.
Our motivation for using such minimal model was two-fold. First of all, it greatly simplifies the analysis, in particular allowing us to use the results of \cite{Barvinsky:1981rw}.
Furthermore, it provides the simplest model in which the Immirzi parameter acquires a physical role. One could consider more general non-minimal couplings, like in \cite{Mercuri:2006um, Alexandrov:2008iy}, and hide the dependence on the Immirzi parameter by adding a redundancy of couplings (see footnote \ref{footMercuri}), but deciding which model fits better to observational criteria goes beyond our present analysis and scope. We limit ourselves only to the following observations about other models. 
First of all, fine-tuned couplings might not be preserved by radiative corrections.
In particular, the coupling proposed in  \cite{Mercuri:2006um} effectively corresponds to Einstein-Cartan theory plus the Nieh-Yan invariant, minimally coupled to fermions. Clearly, given the topological nature of the Nieh-Yan term, the resulting interaction \eqref{Sint} is in this case the same as for pure Einstein-Cartan ($\th=0$ and $\g\to\infty$). However we have shown that such a theory is not stable under renormalization. Secondly, a model with $\th\neq 0$ is very likely to lead to quadratic divergences non-renormalizable within the Holst action, and one needs to consider a more general bare action with all dimension-two invariants.
This generalization, as well as the extension of the calculations to Dirac spinors, are interesting lines of research that we hope to come back to in future work.

\section{Summary and conclusions}
\label{Sec:Conclusions}

We have studied perturbative quantum gravity in the first-order formalism.
We started with the bare Holst action with zero cosmological constant, and quantized the theory around an invertible background metric.
As in the standard second-order metric formalism, the result is a quantum effective action which is on-shell finite for pure gravity, and non-renormalizable in the presence of matter. The quantum theory should then be seen as an effective field theory \cite{Donoghue}.

Our main interest was the effect of quantization on the Immirzi parameter $\g$.
At the classical level, as we recalled and explained, $\g$ plays no role for pure gravity: the equations of motion are independent of it.
Adding fermions with a (possibly non-minimal) Dirac action, we have a source for torsion, and the Immirzi parameter enters non-trivially in the equations of motion. In particular, one finds an effective four-fermion interaction proportional to $\g^2/(\g^2-1)$, working with Euclidean signature. The coefficient interpolates between the interaction in the second-order formalism at $\g=0$, and the pure Einstein-Cartan case at $\g=\infty$. It is singular at $\g^2=1$, where the system collapses to gravity in self-dual variables and no fermions. This simply signals the inconsistency of the Dirac action to couple fermions to the special case of gravity in self-dual variables, for which an alternative fermionic action exists \cite{Romano}. 

At the quantum level, we confirmed the inessential character of the Immirzi parameter in the case of pure gravity. We nevertheless explored the possibility of defining an off-shell running of $\g$, and saw it matching the classical expectation of a privileged role for $\g=0$ and $\g=\infty$. These turn out to be fixed points, respectively IR and UV attractive.

In the presence of fermions, we obtained a physical on-shell running of the Immirzi parameter. 
The running is driven by the divergences associated with a four-fermion interaction generated by radiative corrections.
We computed the beta function, and observed that neither $\g=0$ nor $\g=\infty$ are fixed points of the theory. 
An immediate consequence of these results is that fine-tuned bare actions are not stable under renormalization. These include the second order tetrad action, pure Einstein-Cartan, as well as the coupling to the sole Nieh-Yan invariant considered in \cite{Mercuri:2006um}.
We find instead a UV fixed point at $\g^2=1$, but this leads to a diverging effective four-fermion coupling, and hence it is outside of the validity of perturbation theory. We did not investigate the issue of radiative stability of the gravity-fermions coupling in self-dual variables.

Our results can be extended in a number of directions. Here we considered only Majorana spinors, which allowed us to directly adapt the results of \cite{Barvinsky:1981rw}. One advantage of Majorana spinors is the availability of a spinor-dependent gauge fixing which preserves the minimal form of the gravitational Hessian. Extending the calculations to Dirac spinors requires then dealing with operators in non-minimal forms, or finding a new gauge condition with the same property.

It would be also interesting to start with a more general bare action, which is restricted to invertible tetrads, and includes the full set of invariants \Ref{dim2}. This should allow us to consider the most general non-minimal coupling of fermions, and still be able to renormalize the quadratic divergences. 
A somewhat opposite direction of investigation, is to take more seriously the contribution of degenerate tetrads. 
Throughout the paper, in both the vacuum case and the coupling to fermions, we did not truly consider the contributions of non-invertible tetrads: we worked in perturbation theory around an invertible background, thus effectively neglecting  the contribution of degenerate tetrads. On the other hand, the Holst action is defined also for degenerate tetrads, and so can be the Dirac action on curved spacetime \Ref{Sfermi}, by simply noticing that we can write $e e^\m_I = \f16 \eps^{\m\n\r\s} \eps_I{}^{JKL} e_\n^J e_\r^K e_\s^L$, and thus the inverse tetrad is never needed. 
Solutions with non-invertible tetrads and non-zero torsion are known, and these could give interesting contributions to the path integral, and change our results at the level of both pure gravity and fermion coupling.
The real obstacle in the exploration of more general backgrounds, and in particular of the so-called symmetric phase $\la e\ra=\la \om\ra=0$, comes from the absence of a quadratic term in the action, as it has been well known since long time.
New insight in this direction might come from the Plebanski formulation of gravity, with its relation to topological theories \cite{Plebanski, Reisenberger}.

Finally, we stress that in both the vacuum and the coupled cases, the running of the Immirzi parameter is driven by quadratic divergences.  Quadratic divergences have recently received a lot of attention \cite{Robinson,Toms}, but their physical relevance is still debated (e.g. \cite{Anber}). In particular, a subtle issue with quadratic divergences is that even on-shell they might carry some residual dependence on the gauge-fixing condition (although not necessarily so, see \cite{Niedermaier}). A framework to improve the effective action and ensure gauge-condition independence is that of the Vilkovisky-DeWitt effective action \cite{Vilko} (and it is indeed the one used in \cite{Toms}), and it would be interesting to adapt it to the present case.
The inclusion of a cosmological constant term would also be interesting in this respect, as it would probably add a logarithmic contribution to the running of the Immirzi parameter.

\subsection*{Acknowledgements}
Simone gratefully acknowledges support from the European Science Foundation (ESF) through the activity ``Quantum Geometry and Quantum Gravity'', and wishes to thank the Albert Einstein Institute for hospitality during part of this work.

\appendix
\section{Notation, conventions and useful formulas}
\label{App:Conventions}

Throughout the paper, we work with Euclidean signature and local gauge group SO(4).
We define the Levi-Civita symbol such that $\eps^{\m\n\r\s} = 1$ for $(\m\n\r\s)=(0123)$, from which the tetrad determinant reads
$$
e = \f1{4!} \eps_{IJKL}\eps^{\m\n\r\s} e_\m^I e_\n^J e^\r_K e^\s_L.
$$
We use the same notation for the covariant density weight
$\eps_{\m\n\r\s} = g_{\m\a} g_{\n\b} g_{\r\g} g_{\s\d} \eps^{\a\b\g\d}$
(thus $\eps_{\m\n\r\s} = e^2$ for $(\m\n\r\s)=(0123)$). Some useful formulas are
\be
\eps^{\m\n\r\s}e_\m^I e_\n^J = e\, \eps^{IJKL} e^\r_K e^\s_L, \qquad
\f{1}{4e^2} \eps_{\m\n\a\b}\eps^{\r\s\a\b} = \d_{\m\n}^{\r\s} \equiv \f12 (\d_\m^\r \d_\n^\s - \d_\m^\s \d_\n^\r).
\ee

The curvature and torsion are defined as
\be
F(\om) = d \om + \f12 [\om, \om], \qquad \label{defT}
T(e,\om) = d_\om e,
\ee
or in components
\bea
F(\om)^{IJ}_{\m\n} &=&  \p_\m \om_\n^{IJ} - \p_\n \om_\m^{IJ} + \om_{\m K}^I \om_\n^{KJ} - \om_{\n K}^I \om_\m^{KJ},
\\ 
T(e,\om)_{\m\n}^I &=& \p_\m e_\n^I - \p_\n e_\m^I + \om^{IJ}_\m e_{\n J} - \om^{IJ}_\n e_{\m J}.
\eea

From the decomposition $\om_\m^{IJ}=\om_\m^{IJ}(e)+K^{IJ}_\m$, one finds for the curvature
\be
e_{\m K} e_{\n L} F^{KL}_{\r\s}\big(\om(e)+K\big) = R_{\m\n\r\s}(e) +2\na_{[\r}K_{\s]\m\n} + K_{\r \m\l} K_{\s}{}^\l{}_{\n}-K_{\s \m\l} K_{\r}{}^\l{}_{\n}, 
\ee
where $    {R^\rho}_{\s\mu\nu} = \p_\mu\Gamma^\rho_{\s\nu} - \p_\n\Gamma^\rho_{\s\m} + \Gamma^\rho_{\l\m}\Gamma^\l_{\s\nu} - \Gamma^\rho_{\l\n}\Gamma^\lambda_{\s\m}$ is the Riemann tensor of $e$, related to the Levi-Civita connection $\na_\m$ of $e$ by
\be\label{riemanndef}
[\na_\m,\na_\n]f^\r = R^\r{}_{\s\m\n} f^\s.
\ee
For the torsion,
\be
e^\r_I \, T^I{}_{\m\n}\big(e,\om(e)+K\big) = -2 K_{[\m\n]}{}^\r,
\ee
that is
\be
K_{\m\n}{}^\r = \f12\left(T_{\m\n}{}^\r - T^\r{}_{\m\n} + T_{\n\m}{}^\r \right).
\ee

The connection and contorsion can be decomposed in irreducible representations as in \Ref{omreps} of the main text.
For completeness, we report here the explicit decomposition obtained through the orthogonal projectors 
$\bar P$, $\check{P}$ and $\hat{P}$ defined in \Ref{Preps}. For a field $w$ in \Ref{omreps}, we have
\be
w_{A,BC} = \wb_{A,BC} + \f23 \d_{A[C} \ws_{B]} + \eps_{ABCD} \wh^{D},
\ee
where the irreducible components $\wb$, $\ws$ and $\wh$ satisfy
\be
\wb_{A,BC} \d^{AB} = \eps^{ABCD} \wb_{A,BC}  = 0, \qquad \ws_B = \d^{AC} w_{A,BC},  
\qquad \wh^{D} =  \f16 \eps^{ABCD} w_{A,BC}.
\ee

Using this decomposition and \Ref{M22} one computes
\bea
w^{A,BC} M_{22}{}_{A,BC}^{I,JK} w_{I,JK} &=& \wb^{A,BC} P^{I,JK}_{A,BC} \wb_{I,JK} 
  -\f43 \ws_A \ws^A - 12 \wh_A \wh^A +\f{8}{\g} \ws_A \wh^A \nn \\
&=& w^{A,BC}  \left[ \Big(\Pb - 2  \Ps- 2\Ph \Big) P \right]_{A,BC}^{I,JK} w_{I,JK},
\eea
which is used in the main text.
In going from the first to the second line we used $\wb_{A,BC} = \wb_{B,AC}+\wb_{C,BA}$, a simple consequence of the symmetries of $\wb_{A,BC}$,
and the commutativity property $[\Ps+\Ph , P_\pm ] = 0.$




\end{document}